\documentclass{epl2} 
\usepackage{color}
\usepackage{soul}
\setstcolor{red}
\sethlcolor{yellow}
\begin{document}
\title{Dynamics of pulled desorption with effects of excluded volume interaction:
The $p$-Laplacian diffusion equation  and its exact solution}
\shorttitle{Desorption of pulled polymer}

\author{K.L. Sebastian \inst{1,2} \and V.G. Rostiashvili\inst{1} \and T.A. Vilgis \inst{1}}
\shortauthor{K.L. Sebastian \etal}

\institute{                    
  \inst{1} Max Planck Institute for Polymer Research, 10 Ackermannweg, 55128 Mainz, Germany\\
  \inst{2} Department of Inorganic and Physical Chemistry, Indian Institute of Science, Bangalore 560012, India
}


\abstract{We analyze the dynamics of desorption of a polymer molecule which
is pulled at one of its ends with force $f$, trying to desorb it.
We assume a monomer to desorb when the pulling force on it exceeds
a critical value $f_{c}$. We formulate an equation for the average
position of the $n^{th}$ monomer, which takes into account excluded
volume interaction through the blob-picture of a polymer under external
constraints. The approach leads to a diffusion equation with a $p$-Laplacian
for the propagation of the stretching along the chain. This has to
be solved subject to a moving boundary condition. Interestingly, within
this approach, the problem can be solved exactly in the trumpet, stem-flower and stem regimes. In the trumpet regime, we get
$\tau=\tau_{0}n_d^{2}$ where $n_d$ is the number of monomers that have
desorbed at the time $\tau$. $\tau_{0}$ is known only numerically,
but for $f$ close to $f_{c}$, it is found to be $\tau_{0}\sim f_c/(f^{2/3}-f_{c}^{2/3})$. If one used simple Rouse dynamics, this result changes to {\normalsize $\tau\sim f_c n_d^2/\left(f-f_{c}\right)$.} In the other regimes too, one can find exact solution, and interestingly, in all regimes $\tau \sim n_d^2$.}
\maketitle
\section{Introduction}

Single molecule experiments, in which one exerts an external force
on a long chain molecule have become ubiquitious \cite{Dudko2008}.
These experiments provide very interesting information on the structure
and dynamics of long chain molecules. Mechanical force can be used
to unzip the double stranded DNA, emulating the action of enzymes
that mechanically force the two strands apart during the process of
replication \cite{Bustamante2003}. Force has also been used to desorb
a molecule adsorbed on a surface \cite{MishraEPL2005,Staple2011}.
Motivated by the experiments, the statistical mechanics of an adsorbed
polymer molecules, at one end of which a force is acting trying to
desorb it has been extensively studied \cite{Sebastian1992,BhattacharyaEPJ2009,BhattacharyaMA2009desorb}.
Parallel to these are statistical mechanics of DNA unzipping by force
\cite{Bhattacharjee2000}. In spite of a large amount of activity
on the equilibrium statistical mechanics of these systems, there have
been only few studies of the dynamics. The earliest study of the dynamics
of such a system seems to be that of Sebastian \cite{SebastianPRE2000},
who investigated the Rouse dynamics of a long chain molecule pulled
out from a potential well. It was found that the escape of the polymer
from the well happened if the pulling force $f$ exceeded a critical
value $f_{c}$ and that the time needed for escape of $n_d$ monomers
scaled as $n_d^{2}/(f-f_{c})$. Similar results were obtained by Marenduzo
et. al. who did simulations\cite{Marenduzzo}. There have also been
simulations of the desorption of a pulled polymer \cite{Milchevdesorption}.
However, there has not been any analytic theory of the dynamics of
this process, with excluded volume interactions included. In this
letter, we develop such a theory based on the equations originally
formulated by Brochard-Wyart et al \cite{Brochard-WyartEPL1994}.
The approach leads to a diffusion type of equation for the propagation
of stretching along the polymer. As it accounts for excluded volume
interactions, the equation is non-linear and in the mathematics literature
it is referred to as diffusion equation with $p$-Laplacian \cite{SimsenNA2009}
In the case of pulled desorption, one has to solve this non-linear
equation with a moving boundary condition. Interestingly, an exact
solution can be found for this problem for all the limits of interest.  In the trumpet reigme, the time
for desorption of $n_d$ monomers $\tau$ scales $f_c n_d^{2}/\left(f^{2/3}-f_{c}^{2/3}\right)$, very similar to what was found for the much simpler model of escape
from a well by pulling \cite{Sebastian2000}. If one adopted Rouse
dynamics, this changes to $\tau\sim f_c n_d^{2}/(f-f_{c})$.  Surprisingly, in the other regimes too, time scales exactly in the same fashion as in the trumpet regime, though the dependence on the forces is changed.

\section{The Model}

\begin{figure}[h]
\includegraphics[height=5cm]{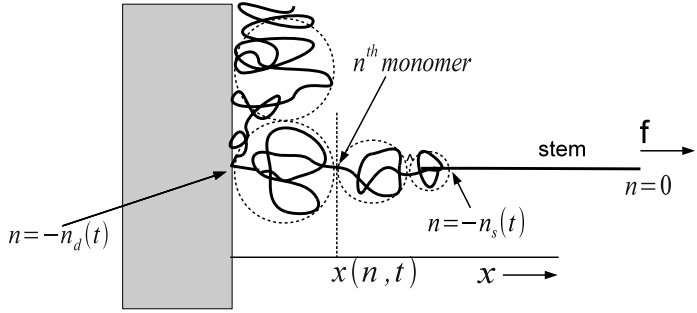}\caption{ The force $f$ is applied in the positive direction,  to the monomer at the end with $n=0$, trying to desorb it.  Desorption occurs if the force at the desorbing monomer (at $n=-n_d(t)$) exceeds $f_c$.  $x(n,t)$ denotes the average position of the $n^{th}$ monomer at the time $t$. The figure shows the stem-flower regime.  In the trumpet regime, the stem is absent (i.e. $n_s(t)=0$). }
\label{Flo:pulled-desorption} 
\end{figure}

We think of a polymer of $N$ monomers, with excluded volume interactions.
The unadsorbed molecule would have a Flory radius $R_{F}\sim aN^{\nu}$,
where $a$ is the size of a monomer. We shall label the monomer position
along the countour length of the monomers with $n$ which varies from
$-N$ to $0$. It is convenient to imagine this as a continuous variable.
The solid surface on which the polymer is adsorbed is located at $x=0$
and at the time $t=0,$ one starts pulling on the end with $n=0$
with a force $f$ in the positive x-direction. If the force on the
monomer at the surface exceeds a critical value $f_{c}$ then the
monomer is assumed to desorb. A simple view of the desorbed part of
the molecule is in terms of the blob-picture introduced by Pincus
\cite{Pincus,PincusMacromolecules76-210,deGennes} and used by Brochard-Wyart et. al \cite{Brochard-Wyart1995,Brochard-WyartEPL1993,Brochard-WyartEPL1994,Brochard-WyartMA1995} and Sakaue \cite{Sakaue2007}
to study the dynamics of pulled polymers.  In this section, we shall discuss what has been referred to as the trumpet regime\cite{Brochard-WyartMA1995}, The other regimes are discussed in a later section.  Let us denote the average
position of the $n^{th}$monomer in the direction of pulling by $x(n,t)$.
As a result of pulling, the end at $n=0$ desorbs and starts moving
as more and more monomers desorb. The desorbed part of the polymer
may be imagined to consists of blobs of (varying) size $R_{b}$ streched
out in the $x-$direction (see Fig. \ref{Flo:pulled-desorption}). Denoting by $\Delta n_{b}$ the number
of blobs of size $R_{b},$we can write 
\begin{equation}
x(n,t)=\sum_{b}R_{b}\Delta n_{b}\label{eq:stretching-out}
\end{equation}
where the sum in Eq. (\ref{eq:stretching-out}) is over all the blobs
to the left of the $n^{th}$ monomer. Let us say that $\Delta n_{b}$
blobs contain $\Delta n$ monomers in total. As a blob of size $R_{b}$
contains $\left(R_{b}/a\right)^{1/\nu}$ monomers, where $\nu$ is
the Flory exponent, $\Delta n$ and $\Delta n_{b}$ are related through
$\Delta n_{b}=\Delta n/\left(R_{b}/a\right)^{1/\nu}$. Hence
$x(n,t)=a^{1/\nu}\sum_{b}R_{b}^{1-1/\nu}\Delta n$. The size of a
blob is related to the force by $R_{b}=k_{B}T/f$, where $k_{B}T$
represents the thermal energy. Thus 
$x(n,t)=a\int_{-n_{d}(t)}^{n}dn' \left (\frac{af(n')}{k_{B}T} \right )^{1/\nu-1}
$,
where $-n_{d}(t)$ 
labels the last monomer that is desorbed at the time
$t$. 
On differentiating this with respect to $n$, we get 
\begin{equation}
\frac{\partial x(n)}{\partial n}=a\left(\frac{af(n)}{k_{B}T}\right)^{(1-\nu)/\nu}\label{eq:stretching}\end{equation}
This is the equation that we shall use in our analysis and is contained
in the paper by Brochard-Wyart et. al \cite{Brochard-WyartEPL1994}.
Using this we wish to calculate the average time of desorption of
$n^{th}$ monomer.   We now calculate the frictional force acting on the portion of the chain in the interval  $(x,x+\Delta x)$.  Denoting the viscosity of the medium by $\eta$, a blob of size $R_b$ would experience a frictional force $\eta v R_b$ where $v $ is the velocity with which the blob is moving. Hence the total force experienced by that region would be  $\eta v \sum_{b\; \epsilon\; \Delta x}R_b=\eta \frac{\partial x}{\partial t} \Delta x $. Hence the net force acting on the $n^{th}$ monomer is $f-\eta\int_{n}^{0}\frac{\partial x(n,t)}{\partial t}\frac{\partial x(n,t)}{\partial n}dn$. Using this in Eq. (\ref{eq:stretching}), we get the equation that describes the building up of tension along the chain on the
right hand side of the wall
\cite{Brochard-WyartEPL1994}.
\begin{equation}
\frac{\partial x}{\partial n}=a\left[\frac{a}{k_{B}T}\left(f-\int_{n}^{0}\eta\frac{\partial x}{\partial t}\frac{\partial x}{\partial n}dn\right)\right]^{(1-\nu)/\nu}\label{eq:blob-formation}
\end{equation}
In Eq. (\ref{eq:blob-formation}),
$-n_{d}(t)<n<0$ where $n$ denotes any particular monomer on the
desorbed part of the polymer.  It may be noted that in the Rouse model, the analogue of Eq. (\ref{eq:stretching})
would be linear in the force and may be obtained by putting $\nu=1/2$.  Further, each monomer experiences a force $\eta a v$  (free-draining regime).  Hence in the Rouse case, instead of Eq. (\ref{eq:blob-formation}) we have
\begin{equation}
\frac{\partial x}{\partial n}=a\left[\frac{a}{k_{B}T}\left(f-\int_{n}^{0}\eta a\frac{\partial x(n,t)}{\partial t}dn\right)\right]. \label{eq:Rouse-stretching}  
\end{equation}
\section{The Diffusion Equation with $p$-Laplacian }
  It is convenient to use dimensionless variables (a tilde on top of
the variable indicates that it is dimensionless) $\tilde{x}=x/\xi$,
$\tilde{n}=n/g$, $\tilde{t}=t/\tau_{b}$ with $\xi=k_{B}T/f$, $g=(\xi/a)^{1/\nu}$,
and $\tau_{b}=\eta\xi^{3}/k_{B}T$.  This leads to 
\begin{equation}
\frac{\partial\tilde{x}}{\partial\tilde{n}}=\left[\left(1-\int_{\tilde{n}}^{0}\frac{\partial\tilde{x}}{\partial\tilde{t}}\frac{\partial\tilde{x}}{\partial\tilde{n}}d\tilde{n}\right)\right]^{(1-\nu)/\nu}.\label{eq:dimensionlessequation}
\end{equation}
Differentiating Eq. (\ref{eq:dimensionlessequation}) gives $
\frac{\partial^{2}\tilde{x}}{\partial\tilde{n}^{2}}=\frac{1-\nu}{\nu}\left(\frac{\partial\tilde{x}}{\partial\tilde{n}}\right)^{\frac{2-3\nu}{1-\nu}}\frac{\partial\tilde{x}}{\partial\tilde{t}}$.
 We shall put $\nu=3/5$ and the equation becomes%
\begin{equation}
\frac{\partial^{2}\tilde{x}}{\partial\tilde{n}^{2}}=\frac{2}{3}\left(\frac{\partial\tilde{x}}{\partial\tilde{n}}\right)^{\frac{1}{2}}\frac{\partial\tilde{x}}{\partial\tilde{t}}.\label{eq:The-Differential-Equation}\end{equation}
 Putting $n=0$ in Eq. (\ref{eq:dimensionlessequation}) leads to 
\begin{equation}
\left(\frac{\partial\tilde{x}}{\partial\tilde{n}}\right)_{n=0}=1.\label{eq:Boundary-Condition-Free-End}\end{equation}
 Further, at the monomer that is in the process of desorbing, the
force has to have the critical value, which we denote by $\tilde{f_{c}}$.
Therefore, at the desorbing end, the boundary condition would be
\begin{equation}
\left(\frac{\partial\tilde{x}}{\partial\tilde{n}}\right)_{n=-\tilde{n}_{d}(t)}=\tilde{f}_{c}^{(1-\nu)/\nu}.\label{eq:Boundary-Condition-Desorbing-End}\end{equation}
In terms of dimensioned variables, $\tilde{f}_{c}=f_{c}/f$.  That is,
 it is the ratio of the critical force to the actual force exerted,
and in order to have desorption, $\tilde{f}_{c}<1$. The initial condition
has all the monomers adsorbed.  Thus
\begin{equation}
\tilde{x}(\tilde{n},0)=0.\label{eq:Initial-Condition}
\end{equation}
The equation (\ref{eq:The-Differential-Equation}) is a diffusion
equation with a $p-$Laplacian, with $p=3/2$, which means fast diffusion
in regions where $\frac{\partial\tilde{x}}{\partial\tilde{n}}$ is
small - this means that the tension spreads rapidly in regions where
there is little tension %
\footnote{An equation of the form $\frac{\partial c}{\partial t}=D\frac{\partial}{\partial x}(\frac{\partial c}{\partial x})^{p-2}\frac{\partial c}{\partial x}$
is known as a diffusion equation with $p$-Laplacian. This has a diffusion
coefficient proportional to $(\frac{\partial c}{\partial x})^{p-2}$(
see reference \cite{SimsenNA2009}). %
}. 

In the case of Rouse dynamics, Eq. (\ref{eq:Rouse-stretching}) would lead to the diffusion equation $\frac{\partial x}{\partial t}=\frac{k_{B}T}{\eta a^{3}}\frac{\partial^{2}x}{\partial n^{2}}$.
Thus with, or without the excluded volume interaction, the stretching
of the polymer obeys a diffusion equation, and not a wave equation
for which the disturbances propagate as pulses. Therefore, any movement
of tension along the chain is diffusive. It is interesting to note
that excluded volume interaction speeds up the diffusive motion (see
ref {[}19{]}).

\section{The Solution}

As time passes, the polymer desorbs. We shall look for solutions of
the form $\tilde{x}(\tilde{n},\tilde{t})=\tilde{t}^{\gamma}h(\tilde{u})$,
$\tilde{u}=\frac{\tilde{n}+n_{d}(t)}{\tilde{t}^{\beta}}$ and $\tilde{n}_{d}(\tilde{t})=\tilde{v}_{d}\tilde{t}^{\alpha}$.
We take $\tilde{v}_{d}$ to be a constant, to be determined. We need
to choose values for the unknowns $\alpha,\beta,\gamma$ and $\tilde{v}_{d}$.
The boundary condition at the free end now reads
\begin{equation}
\left(\frac{\partial\tilde{x}}{\partial\tilde{n}}\right)_{\tilde{n}=0}=\tilde{t}^{\gamma-\beta}h'(\tilde{v_{d}})=1,\end{equation}
 which can be satisfied by putting $\gamma=\beta$ and imposing the
condition 
\begin{equation}
h'(\tilde{v}_{d})=1.\label{eq:Free-End-Condition-on-f}
\end{equation}
 Putting the form $\tilde{x}(\tilde{n},\tilde{t})=\tilde{t}^{\beta}h(u)$
into Eq. (\ref{eq:The-Differential-Equation}) leads to an equation
for $h(u)$ that has only the variable $u$ in it if we choose $\alpha=\beta=1/2$.
The resultant equation is \begin{equation}
3h''(u)+\sqrt{h'(u)}\left((u-\tilde{v}_{d})h'(u)-h(u)\right)=0\label{eq:Equation-for-f(u)}. \end{equation}
 In addition, the conditions (\ref{eq:Boundary-Condition-Desorbing-End})
and (\ref{eq:Initial-Condition}) give \begin{equation}
h'(0)=\tilde{f}_{c}^{2/3},\label{eq:Desorbing-End-Boundary-Condition}\end{equation} and
 \begin{equation}
h(0)=0.\label{eq:f(0)condition}\end{equation}
Eq. (\ref{eq:Equation-for-f(u)}) has to be solved subject to the
initial conditions of (\ref{eq:Desorbing-End-Boundary-Condition})
and Eq. (\ref{eq:f(0)condition}). Further, the value of $\tilde{v}_{d}$
is uniquely determined by the condition of Eq. (\ref{eq:Free-End-Condition-on-f}).
The numerically obtained $\tilde{x}(\tilde{n},\tilde{t})$ is plotted
against $-\tilde{n}$ for different values of $\tilde{t}$ in Fig.~\ref{Flo:plot-of-x(n,t)}a.
\begin{figure}
\includegraphics[width=\linewidth]{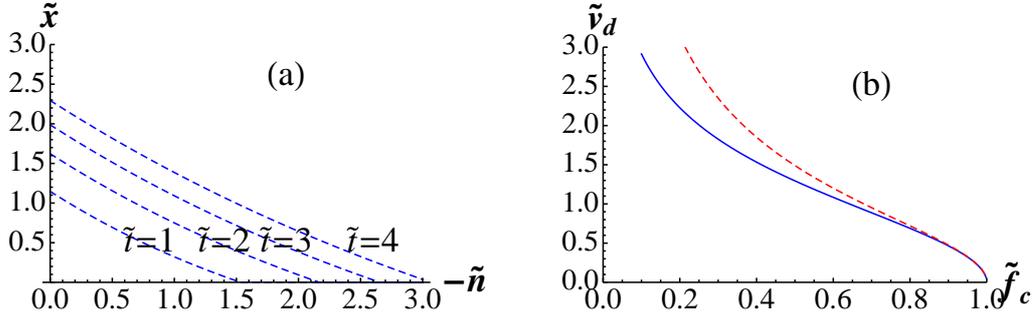}\caption{(a) Plot of $\tilde{x}(\tilde{n},\tilde{t})$ against $-\tilde{n}$ for
different values of $\tilde{t}$. (b) $\tilde{v}_{d}$ against $\tilde{f}_{c}$. The full curve is the result
of numerical calculations while the dashed curve is the result of
using Eq. (\ref{eq:Approximate-vd}). }
\label{Flo:plot-of-x(n,t)}
\end{figure}
One can calculate the position of the pulled end of the polymer as
a function of time, to get $\tilde{x}(0,\tilde{t})=\sqrt{t}h(\tilde{v}_{d})$.
In Fig.~\ref{Flo:plot-of-x(n,t)}b  we give a plot of $\tilde{v}_{d}$
against the dimensionless force $\tilde{f}_{c}$. We now analyze the behavior
of $\tilde{v}_{d}$ in the region where the pulling force exceeds
the critical force only by a small amount. That is, $\tilde{f}_{c}\simeq1$.
In this case $\tilde{v}_{d}\ll1$ and one can do a perturbation analysis
of Eq. (\ref{eq:Equation-for-f(u)}). We put $h(u)=h_{0}(u)+\tilde{v}_{d}h_{1}(u)+\tilde{v}_d^2 h_2(u).$
Our aim is to evaluate $h'(\tilde{v}_{d})$ upto terms quadratic in
$\tilde{v}_{d}$. This can be achieved by evaluating $h_{i}(u)$ correct
upto $u^{3-i}$. Doing this we find that $h_{0}(u)=\tilde{f}_{c}^{2/3}u$,
$h_{1}(u)=f_{c}u^{2}/6$ and $h_{2}(u)=0$. Using these to evaluate
$h'(\tilde{v}_{d})$ gives $h'(\tilde{v}_{d})=\tilde{f}_{c}^{2/3}+\tilde{f}_{c}\tilde{v}_{d}^{2}/3$.
This can be used to solve Eq. (\ref{eq:Free-End-Condition-on-f})
to get
\begin{equation}
\tilde{v}_d = \sqrt{   3(1-\tilde{f}_c^{2/3})/ \tilde{f}_c }.
\label{eq:Approximate-vd}
\end{equation}
 Thus the $\tilde{n}$, number of monomers that have desorbed at the
time $\tilde{\tau}$ is given by
\begin{equation}
\tilde{n}_d=\tilde{v}_{d}\tilde{\tau}^{1/2}.\end{equation}
 Defining $\tau_0=\frac{\tau^* (af_c/k_BT)}{3\left ( (af/k_BT)^{2/3} -(af_c/k_BT)^{2/3}) \right )}$ with $\tau^*=\eta a^3/k_BT$,
 this becomes $
\tau=\tau_0 n_d^2$ in terms of dimensioned variables,
 for $f$ close to $f_{c}$.
The case of Rouse dynamics was discussed in \cite{Sebastian2000}.
The result is (written in the notations of this paper)
\begin{equation}
\tau=f_c n_d^2 / \left( 2\zeta (f-f_c) \right )\end{equation}
for $f$ close to $f_{c}$, where $\zeta=k_{B}T/\eta a^{3}.$ This
is very similar to the result  with excluded volume interaction, except
that the dependence on the force is different in the two cases. The
result $\tau\sim n^{2}$ is in agreement with results of simulation,
which found the total time of desorption of the molecule to be proportional
to $N^{2}$\cite{Milchevdesorption}. The above results show that as the force increases, excluded volume decelerates the desorption.  This is a result of the non-linear stretching of the molecule with excluded volume, leading to an effective friction dependent on the pulling force.
\section{Larger forces: the stem and the stem-flower regimes } 
 \subsection{ The Stem Regime ( $f>f_c>k_BT/a$)}
  In this case, the molecule is stretched to the maximum extend, and would form what has been referred to as the stem \cite{Brochard-Wyart1995}. The desorbed part has $n_s(t)$ monomers and a length of $an_s(t)$.  As a result, it obeys $\eta a^2 n_s(t) \dot{n}_s(t)=f-f_c$.  Introducing the dimensionless variables $\bar{n}_s(t)=n_s(t)$, $\bar{f}=(af/k_BT)$, $\bar{f}_c=af_c/k_BT$,  and $\bar{t}=t(k_B T/\eta a^3)$, this may be solved to get $\bar{n}_s(t) = \sqrt{2(\bar{f}-\bar{f}_c)}\sqrt{\bar{t}}$,   
 showing that in this case too, the desorption time scales like $n_s^2$.
 
 \subsection{The Stem Flower Regime ( $f_c<k_BT/a <f$) \cite{Brochard-Wyart1995}} (see Fig \ref{Flo:pulled-desorption}).  The length of the stem region at any time can be found using the results of previous section, if one remembers that at one end of the stem there is a pulling force $f$ and the other end has $k_BT/a$.  Therefore, the stem consists of $\bar{n}_s(\bar{t})=\sqrt{2(\bar{f}-1)}\sqrt{\bar{t}}$ monomers. The dynamics of the flower part is easily analyzed using the $p-$Laplacian equation.  In terms of $\bar{x}=x/a$, it reads $
\frac{\partial^{2}\bar{x}}{\partial\bar{n}^{2}}=\frac{2}{3}\left(\frac{\partial\bar{x}}{\partial\bar{n}}\right)^{\frac{1}{2}}\frac{\partial\tilde{x}}{\partial\tilde{t}}$, and has to be solved subject to the conditions   $ \left ( \frac{\partial \bar{x}}{\partial \bar{n}}\right )_{\bar{n}=-\bar{n}_s(\bar{t})}=1$, 
 $ \left ( \frac{\partial \bar{x}}{\partial \bar{n}}\right )_{\bar{n}=-\bar{n}_d(\bar{t})}=\bar{f}_c^{2/3}$ and $\bar{x}(\bar{n},0)=0$.  As earlier, putting $\bar{x}(\bar{n},\bar{t})=\sqrt{\bar{t}}h(u)$, $ u = (\bar{n}+\bar{n}_d(\bar{t})/\sqrt{\bar{t}}$, and $\bar{n}_d(\bar{t})=v_d \sqrt{\bar{t}}$, leads to exactly the Eq. (\ref{eq:Equation-for-f(u)}), but with the conditions $h(0)=0, h'(0)=\bar{f}_c^{2/3}$ and $h'(v_d-\alpha)=1$ where $\alpha = \sqrt{2(\bar{f}-1)}$.  The procedure of solution is just the same as earlier and the result is that $\tau
 \propto n_d^2$, the proportionality constant being known only numerically. 

 Finally, we discuss the value of $f_c$ which depends on the adsorption regime. For the weak adsorption regime de Gennes {\it et al.}\cite{deGennesPincus1983}   have shown that the adsorption chain can be seen as a string of  adsorption blobs  with the size of each blob  $D \propto a(\epsilon -\epsilon_c)^{-\nu/\phi}$ , where $\phi$ is the crossover exponent. The dimensionless adsorption energy  $\epsilon = \varepsilon/k_BT$ and $\epsilon_c$ is the critical adsorption energy.  The critical force results from the condition that the size of the Pincus blob is equal to that of the adsorption blob, which leads to $f_c = k_B T(\epsilon -\epsilon_c)^{\nu/\phi}/a $. In the opposite limit of the strong adsorption we have $f_c = \varepsilon/a > k_B T/a$.
 
 \section{Summary and Conclusions}

We have analyzed the dynamics of desorption of an adsorbed polymer
as a result of pulling at one of its ends with a force $f$. Our analysis
takes into account the non-linear stretching of the polymer, which is due to the excluded
volume interactions. The resultant $p$-Laplacian diffusion equation
has to be solved subject to a moving boundary condition. Interestingly, exact solutions can be found in all the regimes of interest.  In the trumpet regime  we find that the time $\tau$ required to desorb
$n_d$ monomers is given exactly by $\tau=\tau_{0}n_d^{2}$ with $\tau_{0}=a^{10/3}f_{c}\eta/\left (3(k_{B}T)^{4/3} (f^{2/3}-f_{c}^{2/3})\right ),$
for $f$ close to $f_{c}$. For the simple Rouse model, this gets
modified to $\tau= f_{c}n_d^{2}/\left(2\zeta(f-f_{c})\right)$. In the stem and stem-flower regime too, time has a quadratic dependence on the number of desorbed monomers, though force dependence is different.
\acknowledgements
We are indebted to A. Milchev for useful discussions.  This investigation has been supported by the Deutsche Forshungemeinschaft (DFG) Grant Nr. SFB 625/B4.  K.L. Sebstian thanks the Max Planck Institue for Polymer Research in Mainz, Germany for hospitality during his vist to the Institute.  K.L. Sebastian has financial support through the J.C. Bose fellowship program of Department of Science and Technology, Government of India. 
\bibliographystyle{eplbib} 
\bibliography{pulleddesorption}

\begin{thebibliography}{10}
\expandafter\ifx\csname url\endcsname\relax\def\url#1{\texttt{#1}}\fi

\bibitem{Dudko2008}
\Name{Dudko O.~K., Hummer G. \and Szabo A.} \REVIEW{Proc. of Nat. Acad. Sci.
  }{105}{2008}{15755}.

\bibitem{Bustamante2003}
\Name{Bustamante C., Bryant Z. \and Smith S.~B.} \REVIEW{Nature
  }{421}{2003}{423}.

\bibitem{MishraEPL2005}
\Name{Mishra P., Kumar S. \and Singh Y.} \REVIEW{Europhysics Letters
  }{69}{2005}{102}.

\bibitem{Staple2011}
\Name{Staple D.~B., Geisler M., Hugel T., Kreplak L. \and Kreuzer H.~J.}
  \REVIEW{New Journal of Physics }{13}{2011}{013025}.

\bibitem{Sebastian1992}
\Name{Sebastian K.~L.} \REVIEW{Chemical Physics Letters }{194}{1992}{375}.

\bibitem{BhattacharyaEPJ2009}
\Name{Bhattacharya S., Milchev A., Rostiashvili V. \and Vilgis T.} \REVIEW{Eur.
  Phys. J E }{29}{2009}{285}.

\bibitem{BhattacharyaMA2009desorb}
\Name{Bhattacharya S., Rostiashvili V., Milchev A. \and Vilgis T.~A.}
  \REVIEW{Macromolecules }{42}{2009}{2236}.

\bibitem{Bhattacharjee2000}
\Name{Bhattacharjee S.~M.} \REVIEW{Journal of Physics A }{33}{2000}{L423}.

\bibitem{SebastianPRE2000}
\Name{Paul A.~K. \and Sebastian K.~L.} \REVIEW{Physical Review E
  }{62}{2000}{927}.

\bibitem{Marenduzzo}
\Name{Marenduzzo D., Bhattacharjee S.~M., Maritan A., Orlandini E. \and Seno
  F.} \REVIEW{Physical Review Letters }{88}{2002}{028102}.

\bibitem{Milchevdesorption}
\Name{Milchev A.} \REVIEW{Private Communication }{}{2011}{}.

\bibitem{Brochard-WyartEPL1994}
\Name{Brochard-Wyart F., Hervet H. \and Pincus P.} \REVIEW{Europhysics Letters
  }{26}{1994}{511}.

\bibitem{SimsenNA2009}
\Name{Simsen J. \and Gentile C.~B.} \REVIEW{Nonlinear Analysis
  }{71}{2009}{4609}.

\bibitem{Sebastian2000}
\Name{Sebastian K.~L.} \REVIEW{Physical Reivew E }{62}{2000}{1128}.

\bibitem{Pincus}
\Name{Pincus P.} \REVIEW{Macromolecules }{9}{1976}{386}.

\bibitem{PincusMacromolecules76-210}
\Name{Pincus P.} \REVIEW{Macromolecules }{10}{1976}{210}.

\bibitem{deGennes}
\Name{deGennes P.} \Book{Scaling Concepts in Polymer Physics} (Cornell
  University Press, Ithaca) 1979.

\bibitem{Brochard-Wyart1995}
\Name{Brochard-Wyart F.} \REVIEW{Europhysics Letters }{30}{1995}{387}.

\bibitem{Brochard-WyartEPL1993}
\Name{Brochard-Wyart F.} \REVIEW{Europhysics Letters }{23}{1993}{105}.

\bibitem{Brochard-WyartMA1995}
\Name{Marciano Y. \and Brochard-Wyart F.} \REVIEW{Macromolecules
  }{28}{1995}{985}.

\bibitem{Sakaue2007}
\Name{Sakaue T.} \REVIEW{Physical Review E }{76}{2007}{021083}.

\bibitem{deGennesPincus1983}
\Name{deGennes P. \and Pinus P.} \REVIEW{J.Phys. (France) Lett.
  }{44}{1983}{L241}.

\end{thebibliography}
 \end{document}